\documentclass[runningheads]{llncs}
\usepackage[T1]{fontenc}
\usepackage{graphicx} 
\usepackage[hidelinks]{hyperref}
\usepackage{tipa}
\usepackage{comment}

\begin{document}

\title{bikiDATA: A Python Library to Query and Explore Large-Scale RDF Datasets}

\titlerunning{bikiDATA: A Python Library for Large-Scale RDF Datasets}

\author{Etienne Posthumus\inst{1}\orcidID{0000-0002-0006-7542} \and
Sven Hertling\inst{1,2}\orcidID{0000-0003-0333-5888} \and
Dilek Yargan\inst{1}\orcidID{0000-0001-9618-6740} \and
Harald Sack\inst{1,3}\orcidID{0000-0001-7069-9804}}

\authorrunning{Posthumus et al.}

\institute{
FIZ Karlsruhe -- Leibniz Institute for Information Infrastructure,  
Hermann-von-Helmholtz-Platz 1, 76344 Eggenstein-Leopoldshafen, Germany\\
\email{firstname.lastname@fiz-karlsruhe.de}
\and
Data and Web Science Group, University of Mannheim, Germany \email{firstname.lastname@uni-mannheim.de}
\and
Karlsruhe Institute of Technology,  
Institute of Applied Informatics and Formal Description Methods,  
Kaiserstr. 89, 76133 Karlsruhe, Germany
}

\maketitle

\begin{abstract}
While knowledge graphs offer unparalleled data flexibility, the semantic gap between RDF triples and the native objects used by software engineers remains a significant barrier to entry. Developing knowledge-graph-backed applications typically requires deep expertise in SPARQL and complex data-mapping layers. To lower this threshold, we present bikiDATA: a high-performance storage solution and a Python library engineered for the modern software developer. Unlike traditional wrappers, bikiDATA abstracts the complexities of the RDF data model into a developer-friendly API that feels native to the Python ecosystem. Beyond standard SPARQL support, the system provides a comprehensive suite for production-grade applications, including integrated full-text search, knowledge graph embeddings, and visual similarity search. Already in use in ongoing projects at FIZ Karlsruhe, bikiDATA reduces integration complexity, improves scalability, and enhances query performance.  The source code and executable demo notebook are publicly available at \url{https://github.com/ISE-FIZKarlsruhe/bikidata}. 

\end{abstract}

\section{Introduction}
While Knowledge Graphs (KGs) have matured from academic curiosities into the backbone of global commerce, powering discovery for giants like Google and IKEA \cite{kari2022medium,singhal2012}, a significant barrier remains for the broader software engineering community. Although Python is now the \textit{lingua franca} of AI and data science, the friction of translating between the RDF data model and native application code remains the primary bottleneck to wider adoption. Building a KG-backed service still requires implementing complex SPARQL queries and rigid mapping layers, a process that often feels more like a research task than standard web development. Additionally, the practical challenges of linked data infrastructure also remain. For instance, public SPARQL endpoints, e.g., Wikidata Query Service, impose strict time and result limits, which often cause queries to fail or return incomplete results \cite{pham2025continuation}, and managing large RDF datasets requires substantial computing resources and complex distributed systems \cite{heling2022robust,potter2018dynamic}. 
To close these gaps, we introduce bikiDATA, a high-performance storage format and Python library designed to treat semantic data as a first-class citizen. By prioritizing developer velocity and providing a fully-fledged toolkit for full-text search, semantic embeddings, KG embeddings and visual similarity search, bikiDATA empowers any software engineer to implement large-scale, intelligent applications with the native ease of a modern Python stack.

\section{bikiDATA}\label{biki}
bikiDATA\footnote{\url{https://github.com/ISE-FIZKarlsruhe/bikidata}\newline The project name incorporates a pun derived from the Afrikaans word "bietjie" \textipa{/bik\textsci/} (Dutch: "beetje"), meaning "a small amount", typically pronounced "bee-key" or "biki". The name is used humorously, since the amount of data in Wikidata is anything but small.} is a Python library and storage format designed for querying large RDF datasets. The project was initially developed to enable advanced querying and extraction from Wikidata dumps while avoiding Wikidata’s query timeouts. 

The project was then pivoted to enable advanced analysis of large linked datasets in Python, where once data is imported, the results of the queries written in native Python syntax are returned as Python structures to be directly reused in further queries.

bikiDATA queries are similar to SPARQL queries: they allow filtering the dataset, applying logical constraints over properties and objects, and querying large linked datasets. However, since everything runs directly in Python, these steps are performed in a more accessible and user-friendly fashion. 
It is important to note that bikiDATA is not intended as a replacement for SPARQL, but as an additional modality with which to query RDF data.
Not all SPARQL queries can be mapped to a bikiDATA equivalent because of the simpler query interface.

Being just Python functions, the interface allows users to import the library, execute queries, and manipulate results in Python code without writing SPARQL.
A bikiDATA query uses a filter option containing a list of items to query (similar to SPARQL basic-graph-patterns). Here are examples of the options to retrieve entities from the index dataset:
\begin{itemize}
    \item \texttt{op}: and | or | not
    \item \texttt{p}: <http://ex.com/foo> | id | fts | semantic | regex
    \item \texttt{o}: <http://ex.com/id/1> | random | "some words"   
\end{itemize}

The operator \texttt{op} can be used to combine several operands with a logical combination.
\texttt{p} can be either a property that the user can specify, e.g., an IRI of a property or a predefined constant that handles other query types.
For example, to get the bikiData representation of an entity in the dataset, the query \texttt{p}: id, \texttt{o}: <http://ex.com/id/1> would return the representation of the entity 1.
If the predicate is \texttt{fts}, the \textbf{f}ull\textbf{t}ext \textbf{s}earch is triggered to search for entities that contain the given search text in \texttt{o}.
For \texttt{semantic}, it performs a text-similarity search based on an embedding.
Given an API key from Cohere\footnote{\url{https://cohere.com}}, it uses the embedding model embed-v4.0.
For the \texttt{regex} option, the underlying regex implementation of DuckDB is used to filter for values

Unlike the in-memory implementation of RDFLib\footnote{\url{https://github.com/RDFLib/rdflib}}
, bikiDATA is designed to handle larger datasets while still allowing queries to be executed within practical time limits. HDT~\cite{hdt_one} is another system capable of efficiently managing large RDF datasets. However, compared to bikiDATA, it offers fewer options for advanced filtering, such as full-text and embedding-based search.

A more detailed example of the demo is demonstrated in the Jupyter notebook\footnote{\url{https://colab.research.google.com/github/ISE-FIZKarlsruhe/bikidata/blob/main/docs/eswc_2026/demo.ipynb}}.

\subsection{bikiDATA Architecture}
To enable high-performance queries, bikiDATA utilises DuckDB, a high-per\-for\-mance analytical relational database management system~\cite {raasveldt2019duckdb}. This approach which provides the ability to analyse huge datasets.\footnote{It also leverages state-of-the-art database research from the Centrum Wiskunde \& Informatica (CWI) in Amsterdam and benefits from a substantial developer community.} 
As DuckDB is a library, rather than a server platform, it can be integrated into bikiDATA with minimal configuration for developers. In addition, the database could also be accessed by multiple processes, even in different programming languages.

RDF data is converted from quads, and stored as simple relation tables for triples, IRIs, and literals. To be able to parallelize the import of the datasets and improve performance, a 64-bit integer hash (computed by xxHash\footnote{https://xxhash.com/}) of IRIs and literals is stored in a dedicated triples table.

\subsection{Scalability}

Even though the library is mainly designed to explore and analyze RDF datasets,
we also conducted several smaller experiments to show the scalability of the proposed library.
We compare to state-of-the-art triple stores based on the Sparqloscope benchmark~\cite{bast2025sparqloscope}.
We run our experiments on a server with 500GB RAM and 96 cores.
For comparability, we executed QLever~\cite{bast2017qlever} on the same hardware.

To show the scalability, we selected all queries from the Sparqloscope benchmark that are slower than 3 seconds.
We pre-translate the SPARQL queries to corresponding SQL queries and measure their execution time.
With a cold cache, the query times were a lot faster than state of the art triples stores like QLever. 
Based on these results, the presented library can easily deal with larger datasets and still return to queries in a small amount of time. More detailed results can be found in the corresponding GitHub repository\footnote{\url{https://colab.research.google.com/github/ISE-FIZKarlsruhe/bikidata/blob/main/docs/eswc_2026/dblp.ipynb}}.

\section{Conclusion and Future Work}\label{conc}
bikiDATA not only provides convenient ways of working with very large datasets to understand the shape of the data, identify what is interesting, and reveal discrepancies, but also bridges the gap between RDF knowledge graphs and Python without requiring SPARQL expertise.

Currently in development is adding a "construct" special property to the filter types to enable making use of SPARQL triplestores containing an index over the same data. This allows bikiDATA to become a Python "front-end" to existing triplestores, allowing the easy integration of full-text or semantic queries (via textual embeddings). We are also adding more embedding types to support structural similarity queries (like RDF2Vec) and visual similarity queries (based on CLIP) to provide hybrid query support.

Ultimately, bikiDATA has the power to transform today's expert-only Semantic Web-based development into a powerful, accessible utility for every software engineer by making RDF a first-class citizen in the Python ecosystem.

\bibliographystyle{splncs04}
\bibliography{references.bib}

\end{document}